\documentclass[review]{elsarticle}

\usepackage{color}
\setcitestyle{numbers,square}
\usepackage{epsfig}
\usepackage{color}
\usepackage{times}
\usepackage{mathtools}
\usepackage{amsmath}
\usepackage{amssymb}
\usepackage{indentfirst}
\usepackage{graphicx}
\usepackage{bm}
\usepackage{longtable}
\usepackage{wrapfig}
\usepackage{comment}
\usepackage{geometry}
\biboptions{sort&compress}
\newcommand{\angstrom}{\AA}

\newcommand{\bey}{\begin{eqnarray}}
\newcommand{\eey}{\end{eqnarray}}
\newcommand{\vep}{\varepsilon}

\newcommand{\bec}{\begin{center}}
\newcommand{\eec}{\end{center}}

\usepackage{numcompress}

\makeatletter
\def\ps@pprintTitle{%
  \let\@oddhead\@empty
  \let\@evenhead\@empty
  \def\@oddfoot{}%
  \let\@evenfoot\@oddfoot
}
\makeatother

\begin{document}

\setcounter{tocdepth}{2}
\baselineskip   15pt 
\belowdisplayskip14pt 
\belowdisplayshortskip10pt 
\renewcommand{\thefootnote}{\fnsymbol{footnote}}

\begin{frontmatter}

\title{Unveiling the Atomistic Mechanisms of Shear-Induced LDA$\leftrightarrow$HDA Transformations and Shear Banding in Amorphous Silicon under High Pressures}


\author[mymainaddress]{Hao Chen \corref{mycorrespondingauthor2}}
 \cortext[mycorrespondingauthor2]{Corresponding author}
 \ead{haochen_mech@ujs.edu.cn}
 \author[mythirdaddress]{Valery I. Levitas \corref{mycorrespondingauthor1}}
  \cortext[mycorrespondingauthor1]{Corresponding author}
  \ead{vlevitas@iastate.edu}
  
 \author[mymainaddress]{Tengyi Liu}
 \author[mymainaddress]{Jingyu Lu}
 \author[mymainaddress]{Rui Zhu}
 \author[mymainaddress]{Zhongqiang Zhang}

\address[mymainaddress]{School of Mechanical Engineering, Jiangsu University, Zhenjiang 212013, P.R. China}

\address[mythirdaddress]{Departments of Aerospace and Mechanical Engineering, Iowa State University, Ames, IA 50011, USA}


\begin{abstract}
Phase transformations (PTs) between different amorphous phases under hydrostatic loading and PTs in crystalline materials under severe plastic deformations attract significant attention due to their great fundamental importance and applied potential. However, we are not aware of any publications on plastic shear-induced amorphous-amorphous PTs under pressure. Here, large-scale molecular dynamics simulations of shear deformation under constant pressures of amorphous silicon, PT from low-density-amorphous (LDA) to high-density-amorphous (HDA) Si, and formation of shear bands (SBs) are performed using the state-of-the-art Gaussian Approximation Potential. The simulations reveal that LDA$\leftrightarrow$HDA shear-induced PTs occur simultaneously until reaching steady state. The developed mechanism-based analytical model well describes shear-strain-governed kinetics and steady states at all pressures, independent of shear stresses. Shear reduces the pressure for initiation and completion of LDA$\rightarrow$HDA PT by $4.36$ and $5.10$ GPa, respectively. Without PT at low pressure, shear-banding occurs, which is partially suppressed by PT at higher pressure with  uniform deformation-PT at $9.8$ GPa. Despite the much larger shear and expected fraction of HDA, surprising sharp drop in the HDA atomic fraction within the SB was discovered. In bulk, Si deforms by atomic rearrangement in localized shear transformation zones with high nonaffine displacements, which trigger nucleation of HDA clusters within LDA and, concurrently, of LDA clusters within HDA, without growth and coalescence. In SB, a turbulent-like flow with swirls is revealed, which promotes reverse PT from HDA$\rightarrow$LDA more effectively. Transformation-induced plasticity in amorphous Si is revealed. The findings open up basic research into the mechanisms and kinetics of plastic strain-induced PTs in amorphous materials under high pressure, with numerous important applications.

\end{abstract}


\end{frontmatter}
{\color{black}\footnote{\color{black}HC and VIL contributed to this paper equally}}


After discovering pressure-induced amorphization and LDA$\rightarrow$HDA PT in ice \cite{Mishima-ice-Nature-85}, amorphous-amorphous
PTs were revealed and intensively studied in Ge, 
SiO$_2$, GeO$_2$, TiO$_2$, Rb$_{6.15}$Si$_{46}$ clathrate, Ge$_2$Sb$_2$Te$_5$ phase-change memory material, zeolites, geological, biological, and soft matter systems, etc
\cite{Aasland-a-a-Nature-94,Sun-a-a-Ge-Sb-Te-11PNAS-11,Machon-PMS-14}.
Under hydrostatic pressure, 
amorphous Si (a-Si)  transforms from a covalently bonded, four-fold coordinated LDA phase to a more metallic, highly coordinated HDA phase, as observed experimentally \cite{deb2001pressure} and with molecular dynamics (MD) simulations \cite{deringer2021origins,fan2024microscopic}.
Mechanisms of plastic deformations (shearing) of amorphous materials via nucleation and coalescence of shear transformation zones (STZ) and  formation  of shear bands (SBs) are broadly studied for metallic glasses \cite{Argon-79,Falk-Langer-98,csopu2017atomic,Zhou-BMG-PMS-24} 
and Si \cite{albaret2016mapping}, 
but without amorphous-amorphous PTs. 

Independently,  the effect of plastic shear (using high-pressure torsion) on high-pressure PTs in crystalline materials is under increasing interest \cite{Blank-Estrin-2014,Levitas-MT-23,Review-HPT-JAL-24,Review-ceramics-25,Levitas-PMS-26}, in particular, for Si \cite{Blank-Estrin-2014,Yesudhasetal-NatCom-Si-24},
because it drastically reduces the PT pressure,
accelerates kinetics, changes the PT paths, and leads to new nanostructured  phases not observed under hydrostatic compression.   
These PTs were coined {\color{black} with} the plastic strain-induced PTs under high pressure \cite{Levitas-PRB-04,Levitas-PMS-26}; they require completely different experimental characterization and thermodynamic and kinetic description.

Severe plastic deformations  during high-pressure torsion, in SBs, and other processes can cause amorphization of crystalline 
geological materials \cite{Samae-a-rocks-Nature-21,Idrissi-a-olivine-SB-AM-22}, ceramics \cite{Chen-aBC-Science-03,An-PRL-2014-BC,Blank-Estrin-2014,levitasetal-SiC-12}, and fullerene \cite{Kulnitskiyetal-16}. 
Si-I undergoes amorphization with formation of bands  during nanoindentation   \cite{Kiran-Chapter-15}, compression  of  nanopillars \cite{Heetal-NatNan-16}, shock loading \cite{Zhao-Meyers-AM-16,Meyers-MatTod-21,li-Meyers-2022amorphization},  scratching, and machining \cite{Patten-Cherukuri-Yan-section6-04,Goeletal-15,wang2025systematic}.
In addition to the thermodynamic reasons  caused by the elimination of the energy of dislocations and other defects, amorphization is also considered  an additional carrier for plastic deformation \cite{yan-meyers-review-21shear-amorh,Samae-a-rocks-Nature-21,Idrissi-a-olivine-SB-AM-22,chen2025virtual}.
Formation of amorphous band in perfect crystals have been studied in 
\cite{Hu-a-band-plasticity-NatMat-23,chen2025virtual} for Al$_2$Sm and Al$_3$Sm intermetallics and Si  
\cite{Chen-etal-AM-19,chen2025virtual}.
Shear-induced nanocrystallization in amorphous materials and metallic glasses  \cite{Chen-nanocrystal-SPD-Nature-94},
carbon systems \cite{Shiell-diamond-16,Wong-Bradby-19,Zhang-Crystal-Glass-AdvMat-26}, and in Si and Ge \cite{Zhao-Meyers-AM-16,Haberl-MatTod-25,chen2025virtual} were studied in detail.
However, {\it {\color{black} we} were are not aware of any studies of strain-induced amorphous-amorphous PT for any material system.} 
The basic questions include understanding of: the atomistic mechanisms of plastic deformation of amorphous materials, the SB formation and deformation mechanism in them,  the effect of plastic deformation and SBs on amorphous-to-amorphous PTs and mechanisms of these PTs outside and within the SB, as well as formalizing results in the macroscopic mechanism-based kinetic equation for PT and searching for new phenomena.  
Because of the very small time and spatial scales and reversibility of LDA$\leftrightarrow$HDA PT,  the related atomistic mechanisms  can be currently resolved via atomistic simulations only. In addition to filling a gap in fundamental understanding, this topic has significant applied importance, especially for a-Si studied here, described in Discussion. 



Here, we integrate sciences of severe plastic deformations  of amorphous materials, strain-induced PTs under high pressure, and {\color{black} shear} banding, and extend them to
study the shear deformation under constant pressures in a-Si, LDA$\leftrightarrow$HDA shear-induced PTs, and mutual effects of PT and SBs formation using the state-of-the-art Gaussian Approximation Potential 
\cite{deringer2021origins, fan2024microscopic,bartok2018machine}. 
Shear deformation leads to the shear band, in which a dissipative structure of turbulent-like flow with swirls is revealed, followed by reduction in number of swirls via coalescence with further shearing. Increasing pressure results in the LDA$\rightarrow$HDA PT with increasing atomic fraction $c$ of HDA due to pressure and shear, which in turn suppresses the shear banding,  enabling  the uniform deformation-PT at $9.8$ GPa. 
The simulations reveal that shear-induced LDA$\leftrightarrow$HDA PTs occur simultaneously until reaching steady state. 
Shear reduces the pressure for initiation and completion of LDA$\rightarrow$HDA PT by 4.36 and 5.10 GPa compared to hydrostatic loading, respectively. Atomic mechanism of PT differs fundamentally from that under hydrostatic loading. 
In bulk, Si deforms by atomic rearrangement in localized STZs with high nonaffine displacements, which trigger nucleation of HDA clusters within LDA and, concurrently, of LDA clusters within HDA, without their growth and coalescence. 
Despite the much larger shear and expected $c$ in an SB, a surprisingly sharp drop in $c$ within the SB was discovered. It was found that this anomalous behavior is related to the change in deformation mechanism: swirls in an SB promote reverse PT from HDA$\rightarrow$LDA more effectively than the direct PT. While clear correlation between fields of the nonaffine squared
displacement and the atomic coordination in the bulk is found, such a correlation in a shear band is lacking. There is no correlation between the atomic displacement and coordination number fields, both within and outside the SB. 
We developed a simple, mechanism-based analytical kinetic model that uses accumulated plastic strain rather than time, simultaneous direct-reverse PTs, pressure-dependent driving forces for both PTs, and different yield strengths of the phases. It possesses a steady-state solution depending on pressure. Both non-stationary and stationary solutions accurately describe the MD simulation results across all pressures, allowing us to determine all material parameters. Kinetic equation is also derived for pressure-induced LDA$\leftrightarrow$HDA PTs, and the stationary solution  reproduces the MD simulation results correctly.
Despite the large shear stresses, they surprisingly do not affect the PT kinetics. This happens because, according to atomic mechanisms, shear stresses simultaneously promote both direct and reverse PTs in different regions, resulting in negligible net effects. 
 TRIP due to volume change during amorphous-to-amorphous PT is revealed.
 While for crystalline solids, temperature and/or stress hystereses require temperature and/or stress cycling to accumulate significant TRIP, LDA$\leftrightarrow$HDA PTs occur simultaneously, and each of them produces TRIP.  Consequently, even for relatively low net $c$, TRIP and stress relaxation can be significant due to many back-and-forth events.  An equation for the TRIP-induced deformation rate is proposed that accounts for the above results. 
Our findings open fundamental research on the mechanisms and kinetics of plastic strain-induced PTs in amorphous materials under high pressure, provide deep atomistic and macroscopic insights into the coupled pressure-shear deformation, shear banding, and PT in a-Si, thus offering a crucial theoretical foundation for designing experiments and various engineering applications.
\section*{Results}
\noindent
{\bf Plastic flow without PT in Si at normal pressure} 

\begin{figure} [!htbp]
 \centering
 \includegraphics[width=\textwidth]{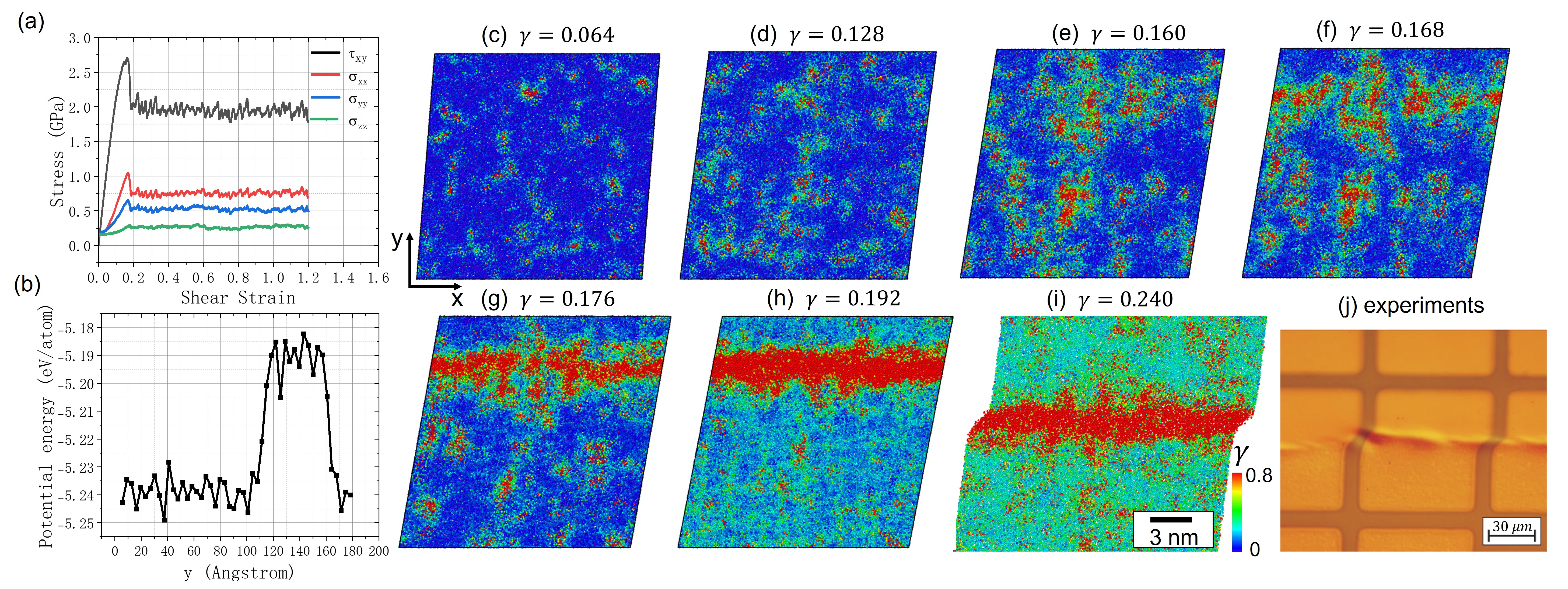}
    \caption{
  {\bf 
Typical shear band formation process in amorphous silicon revealed by MD simulations.}  (a) Stress–strain curves at strain rates of $0.1\ ns^{-1}$ at $10\ K$ for  the whole sample, showing the initial elastic regime, the yielding peak, and subsequent strain softening. (b) Potential energy distribution plotted along the direction perpendicular to the shear band, illustrating an enlargement of potential energy inside the shear band. (c)–(i) Atomic configurations for selected shear strains, shown in the wrapped Eulerian box representation. Starting with random nucleation of the STZ at $\gamma  = 0.064$, density of the STZs increased in (d)-(e)  and progressively coalesce into a well-defined shear band in (f)-(g). (j) Final configuration visualized in the Lagrangian description, and compared with the experimental results in (k) from \cite{hedler2004amorphous}. Reproduced with copyright permission from  {\color{black} Springer Nature}. Local atomic shear strain is color-coded according to \cite{stukowski2009visualization}. 
}
  \label{fig:plotS12}
\end{figure}

Shear stress–strain response in Fig. \ref{fig:plotS12} exhibits an initial elastic regime, followed by a pronounced yielding peak and subsequent sudden stress drop from $2.7\ GPa$ to $2.0\ GPa$, during which the SB formed. Normal stresses exhibit similar behaviour but are much smaller. The potential energy in the SB is higher than in surrounding  (Fig. \ref{fig:plotS12}(b)). Sequential atomic configurations (Fig. \ref{fig:plotS12}(c)–(i)) show multiple nucleations of  clusters of local shear strain (i.e., STZ),  their alignment and coalescence  resulting  into a fully developed SB, qualitatively similar to that in  experiments  in amorphous Si in Fig. \ref{fig:plotS12}(k).

\begin{figure} [!htbp]
 \centering
 \includegraphics[width=0.6\textwidth]{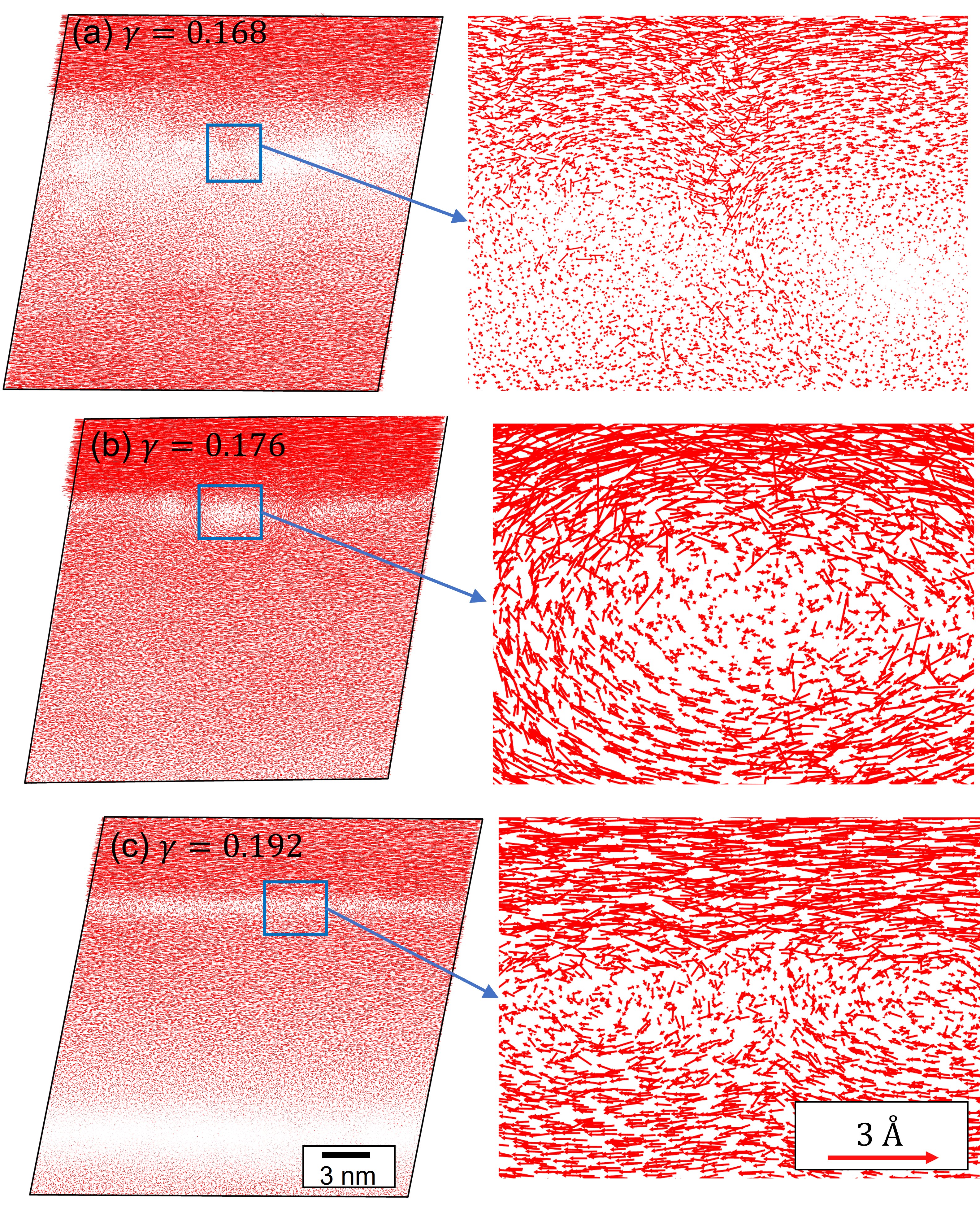}
    \caption{
{\bf
Atomic displacement vector fields illustrating the evolution of shear deformation in a-Si at normal pressure. } (a) 
At $\gamma = 0.168$, individual shear transformation zones begin to coalesce, initiating swirl-like motion. (b) At $\gamma = 0.176$,  two large eddy-like rotational flows are developed.
(d) At $\gamma = 0.192$, a fully developed shear band emerges with  smaller internal swirls. 
}
  \label{fig:plotS13}
\end{figure}

\begin{figure} [!htbp]
 \centering
 \includegraphics[width=0.65\textwidth]{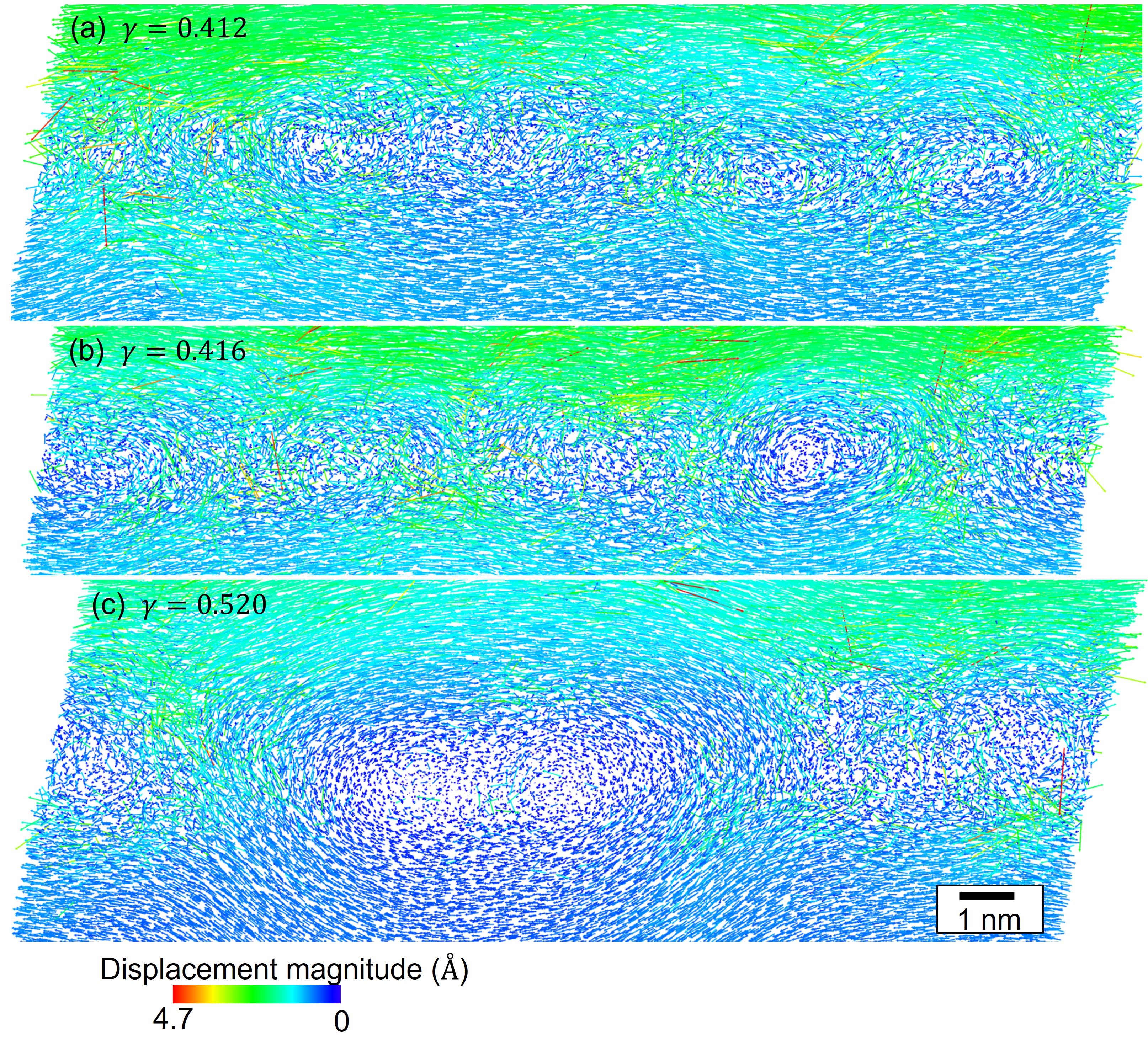}
    \caption{
  {\color{black}  {\bf Atomic displacement vector maps inside the shear band in a-Si.} (a) At $\gamma = 0.412$, five vortex-like swirls with $~3.5\ nm$ spacing are observed. (b) At $\gamma = 0.416$, two swirls merge, leaving four with $~4.4\ nm$ spacing. (c) At $\gamma = 0.520$, further coalescence produces three larger swirls. The evolution reveals a novel mechanism of 
  the reduction of the number of swirls in the band via coalescence, not reported for any material.
}
}
  \label{fig:plotS14}
\end{figure}

\begin{figure} [!htbp]
 \centering
 \includegraphics[width=\textwidth]{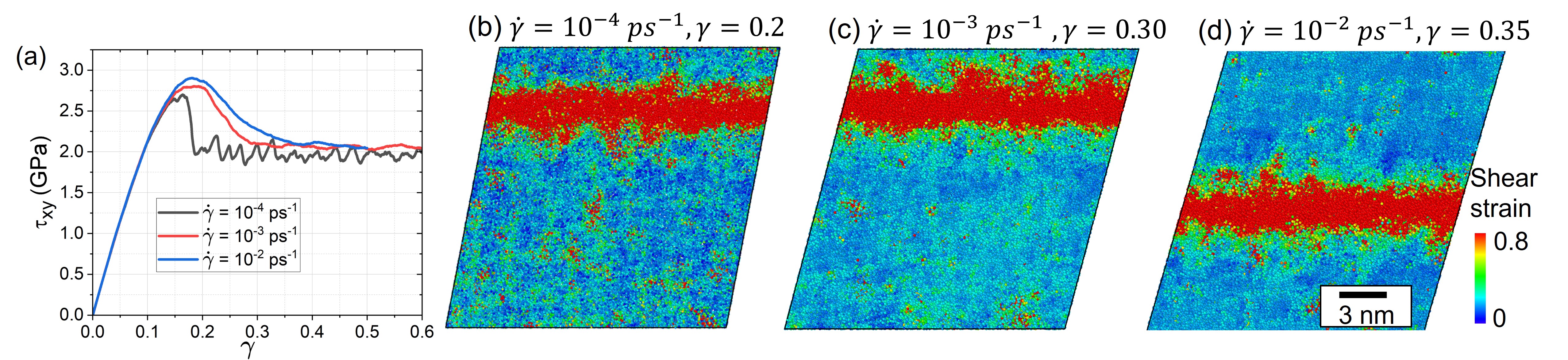}
    \caption{
  {\color{black}  {\bf Strain-rate effect on the mechanical behavior of a-Si.} (a) Stress-strain curves at different strain rates ;  (b)-(d) Shear strain fields with the SBs formed at different strain rates. 
}
}
  \label{fig:plotS17}
\end{figure}

The distribution of the displacement vector field of atoms during shear band formation  in Figs. \ref{fig:plotS13} and \ref{fig:plotS14}, revealed a unique sequence of structural evolution.
At $\gamma = 0.168$, localized STZs begin to coalesce, giving rise to the appearance of swirls,  evolving  into two swirls at $\gamma = 0.176$.
At $\gamma = 0.192$,   a fully developed shear band forms that contains the swirls. 
With increasing shear, internal structure in the entire band in Fig. \ref{fig:plotS14} evolves with coalescence from five  swirls  with an average  spacing of  $\sim$3.5nm at $\gamma = 0.412$ to  four swirls with a spacing of  $\sim$4.4 nm at $\gamma = 0.416$, and three swirls, which are still in the coalescence process 
at $\gamma = 0.520$. 
Such a turbulent-like flow in shear band and formation of evolving vortex-like dissipative structures were not observed in the previous
studies for Si \cite{albaret2016mapping}. 
To some extent, similar vortex structure was found in MD simulations of Cu$_{64}$Zr$_{36}$ metallic glass. However, a shear band was initiated and propagated from the notch; stresses and strains were heterogeneous along the band, strains did not exceed $0.08$, and reduction in number of swirls via their coalescence was not observed. We will show below that formation of swirls changes mechanism and kinetics of strain-induced LDA$\leftrightarrow$HDA PT. 
Increase in the shear rate by one and two orders of magnitude increases and broadens the stress peak and smoothes post-peak stress-strain curve (Fig. \ref{fig:plotS17}). At the lowest strain rate of $10^{-4}\ ps^{-1}$, the stress exhibits a pronounced serrated flow, characteristic of intermittent localized rearrangements and stick-slip dynamics, which is the same as quasi-static loading for  \cite{albaret2020time}. 
Smoothing of the stress curves for higher $\dot\gamma$ suggests that the system has less time to relax between successive atomic rearrangements and is reminiscent of the behavior  in metallic glasses
\cite{greer2013shear, antonaglia2014bulk}. Despite these differences in stress–strain response, SBs  are practically independent of the strain rates (Fig. \ref{fig:plotS17} (b)-(d)). Note that analysis of the pair correlation function $g(r)$  and bond-angle distribution  of atoms within and outside the SB for different strains confirms that the structure remains in the $LDA$ phase without essential PT.


\noindent
{\bf LDA$\leftrightarrow$HDA PT at high pressures}
 
\begin{figure} [!htbp]
 \centering
 \includegraphics[width=\textwidth]{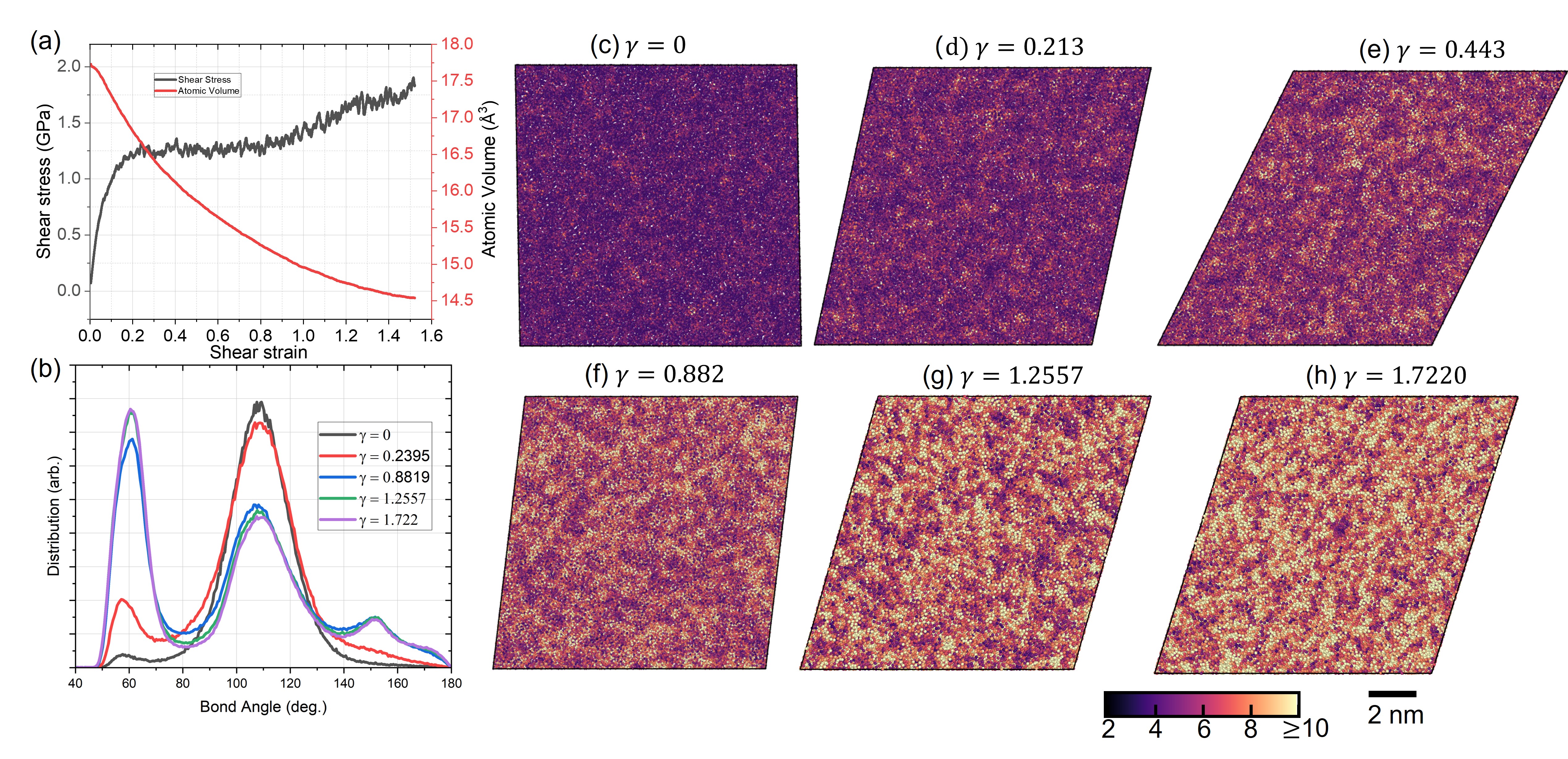}
    \caption{
  {\color{black}  {\bf Mechanical response and structural evolution in the whole a-Si sample under shear straining at   pressure of 9.8 GPa.} (a) Shear stress and atomic volume per atom as functions of $\gamma$; (b) Bond angle distributions  at selected shear strains; (c)–(h) Atomic configurations at increasing shear strains, with atoms colored by coordination number. 
}
}
  \label{fig:plotS19}
\end{figure}

Fig. \ref{fig:plotS19} illustrates the mechanical response and accompanying structural evolution in a-Si under shear at constant pressure of $9.8$ GPa.
Fig. \ref{fig:plotS19}(a) presents the shear stress–strain and atomic volume–shear strain relationships. The shear stress increases rapidly at small strain and then gradually rises without exhibiting a pronounced stress drop, indicating the absence of macroscopic shear localization. Meanwhile, the atomic volume decreases monotonically with increasing shear strain and then reaches a stable value of  $14.5 \angstrom ^3$, corresponding to the atomic volume of HDA silicon \cite{deringer2021origins, fan2024microscopic}, demonstrating a continuous shear-induced  PT process. Fig \ref{fig:plotS19}(b) shows the evolution of the bond angle distribution at selected shear strains. With increasing shear strain, the dominant peak near the tetrahedral angle ($\sim$109.5$^{o}$), characteristic of LDA, progressively weakens, while the peak of smaller bond angles (around $60^{o}$) increases. This redistribution of bond angles reflects a gradual breakdown of the tetrahedral network and the development of more distorted and densely packed local motifs, providing direct structural evidence of a shear-induced LDA$\rightarrow$HDA  PT in Si, similar to that under hydrostatic loading in  \cite{fan2024microscopic}. The corresponding atomic configurations in Fig. \ref{fig:plotS19}(c)–(h) depict the microstructural evolution during shearing. LDA Si corresponds to the coordination number around $4$;  for quantitative analysis of the PT kinetics we accept that coordination numbers $\geq 6$ indicate a HDA  phase \cite{deringer2021origins}. At small shears, the structure is dominated by fourfold-coordinated atoms, consistent with a tetrahedral LDA network. As shear strain increases, the fraction of overcoordinated atoms (five- and sixfold coordinated) grows progressively and becomes spatially homogeneous throughout the sample. No localized SB is observed; instead, the coordination changes are uniformly distributed, indicating a bulk transformation process. At large shear strain, the microstructure is characterized by a high population of overcoordinated atoms and a substantially densified network, corresponding to a HDA phase.

\begin{figure} [!htbp]
 \centering
 \includegraphics[width=\textwidth]{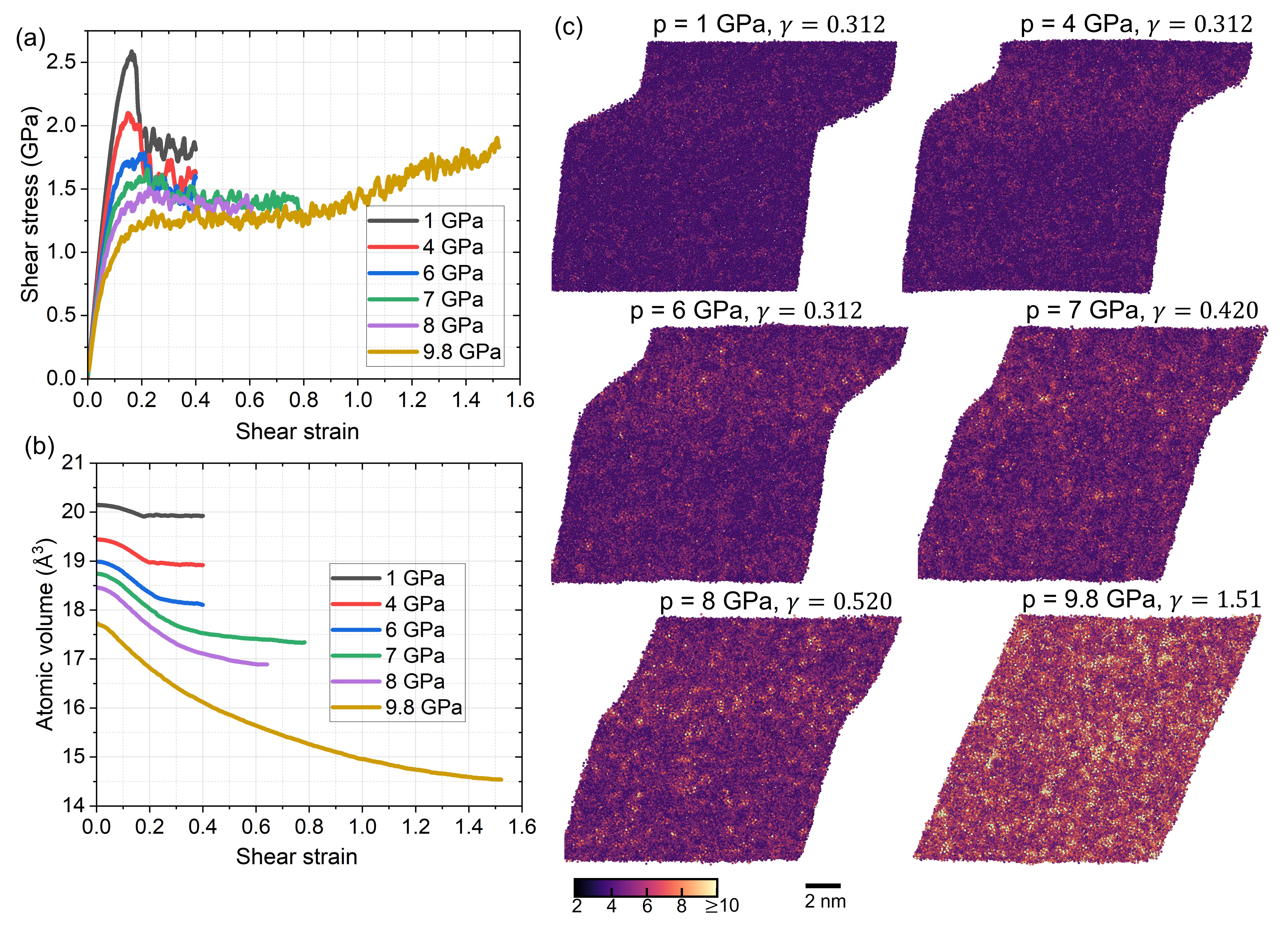}
    \caption{
  {\color{black}  {\bf Mechanical response and structural evolution of a-Si  under shear deformation at different constant  pressures $p$.} (a) Shear stress-shear strain curves for pressures ranging from $1$ to $9.8$ GPa for the whole sample; (b) Atomic volume per atom as a function of shear strain under the same loading conditions. The atomic volume is averaged over the volume outside of the SB (in bulk) for $p<9.8$ GPa and  over the entire volume for  $9.8 GPa$. (c) The actual atomic configurations corresponding to the Lagrangian description colored by atomic coordination number. 
}
}
  \label{fig:plotS18}
\end{figure}

\begin{figure} [!htbp]
 \centering
 \includegraphics[width=\textwidth]{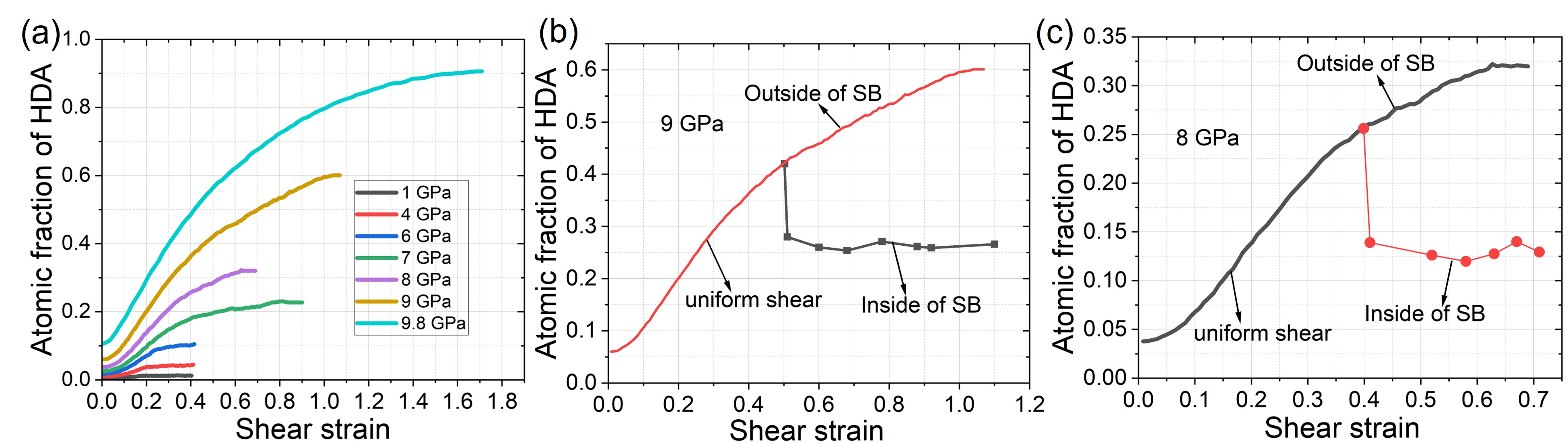}
    \caption{
  {\color{black}  {\bf Kinetics of shear strain-induced LDA$\leftrightarrow$HDA transformation in a-Si under different   pressures.} (a) Atomic fraction of HDA silicon versus shear strain under different pressures shown in the plot. For  $p<9.8 GPa$, the atomic fraction of HDA is averaged over the volume outside of the SB. For  $9.8 GPa$, the atomic fraction of HDA is averaged over the entire volume. 
  (b)-(c) Atomic fraction of HDA silicon versus shear strain at  $9$ GPa (b) and $8$ GPa (c), distinguishing between regions inside and outside the formed SB. Significant sharp drop in HDA atomic fraction within the SB is observed. 
}
}
  \label{fig:plotS28}
\end{figure}

Fig. \ref{fig:plotS18}(a) shows the shear stress–strain curves for constant applied pressures ranging from $1$ to $9.8$ GPa. At low pressures (1 to 8 GPa), shear stress increases rapidly at small strain, reaches a pronounced peak stress, and then exhibits strain softening accompanied by stress fluctuations. This behavior is characteristic of shear localization and the onset of shear banding (Fig. \ref{fig:plotS18}(c)). 
In contrast, at the highest pressure of $9.8$ GPa, the stress increases monotonically without an obvious drop which is  consistent with the absence of macroscopic SB formation. 
The atomic volume in Fig. \ref{fig:plotS18}(b) shows a  decrease  followed by a relatively constant value, consistent with the increase in atomic fraction of the HDA $c$ up to a steady value (Fig. \ref{fig:plotS28}(a)). With an increase in  pressure, the overall atomic volume  significantly reduces due to larger $c$. With increasing shear strain at $9.8$ GPa (for which almost complete PT to HDA  is achieved), the atomic volume continuously decreases from  $17.7$ {\AA}$^3$ to a steady value $14.5$ \AA$^3$,  corresponding to LDA and HDA, respectively \cite{deringer2021origins}.

The  actual atomic configurations corresponding to Lagrangian description are displayed in Fig. \ref{fig:plotS18}(c). At pressures in the range of $1-8$ GPa, the coordination number remained largely as $4$ and a distinct SB emerges as deformation proceeds, characterized by a narrow region of high local shear strain. This band forms through the spatial clustering and coalescence of STZ, similar to that without PT. However, formation of a continuous band occurs at shears significantly larger than that corresponding to the stress peak (Fig. \ref{fig:plotS28}(b) and (c)). In contrast, at $9.8$ GPa, significant LDA$\rightarrow$HDA PT occurs and no continuous SB is observed even at a shear strain exceeding $1.5$. 

\noindent
{\bf Transformation-induced plasticity (TRIP) phenomenon for amorphous-to-amorphous PT}

Increasing deviation from the linear behavior and reduction in the peak stress with increasing pressure is caused by increasing atomic fraction of HDA phase $c$ (see $c$ at zero plastic strain in Fig. \ref{fig:plotS28}(a)).  Thus, LDA$\rightarrow$HDA PT represents an effective mechanism of shear stress relaxation. This happens (a) because of  an additional inelastic shear strain due to atomic rearrangement during the PT and (b) due to volume reduction causing significant  internal stresses that promotes plasticity. These mechanisms represent the TRIP, well known for crystalline solids but never discussed for amorphous materials.  The second mechanisms is  quite 
similar to the Greenwood–Johnson mechanism of TRIP, while the first one substitutes the  Magee  mechanisms due to preferred orientation of different martensitic variants \cite{Fischer-etal-00}. As we will discuss below, there is simultaneous direct and reverse 
LDA$\leftrightarrow$HDA PT at different location of the sample, and each of them produces TRIP. Consequently, even for relatively low resultant $c$, TRIP and stress relaxation can be significant due to large number of back-and-forth events.  
At the critical pressure near $9.8$ GPa, TRIP completely eliminates stress peak and strain softening, and therefore initiation shear banding.
Since HDA is stronger than LDA,  shear banding is suppressed during the entire PT. TRIP, as an additional mechanism of stress relaxation, also explains a plateau in the stress-strain curve for pressures in the range of 7-9.8 GPa despite the increase in atomic fraction of the stronger HDA Si for $\gamma \leq 0.8$.  Strain hardening is observed at 9.8 GPa at  $\gamma \geq 0.8$ only due to significant slowdown of the PT kinetics (Fig. \ref{fig:plotS28}(a)) and probably due to increase in the coordination number within HDA (i.e., transformation to very HDA). 
We will not distinguish between HDA and very HDA phases \cite{deringer2021origins, fan2024microscopic}, because due to luck of strict definitions, it would be difficult to develop a quantitative kinetics presented below. Qualitatively, at 9.8 GPa, HDA is observed at small shear and very HDA at large shears (Fig. \ref{fig:plotS19}).  For lower pressures and smaller shears, PT to HDA  mostly occurs   
(Fig. \ref{fig:plotS18}).

\noindent 
{\bf PT kinetic and atomistic mechanism within and outside the shear band}

 Fig. \ref{fig:plotS28}(a) illustrates the evolution of the atomic fraction of HDA Si $c$ as versus applied shear strain for different constant  pressures. Increase of both pressure and shear leads to the increase in $c$. Pressure accelerates  the kinetic by increasing the thermodynamic driving force for PT due to reduction in volume. Shear play a role of a time-like parameters leading {to the increase in number of nucleation sites, i.e., STZs}. All curves tend to or end with plateaus, i.e., steady  atomic fraction of HDA silicon $c_s$, which increases with increasing pressure.  
 At the highest simulated pressure of $9.8$ GPa and uniform deformation ( Fig. \ref{fig:plotS18}), the HDA atomic fraction rapidly increases with shear strain, approaching a near-complete transformation (close to 90$\%$).  In contrast, at lower pressures in the range of  $1-9$ GPa, the transformation to HDA is only partial and cannot prevent the SB formation. Figs. \ref{fig:plotS28} (b)-(c) provide a more localized analysis of the HDA transformation at  pressures of $9$  and $8$ GPa  by differentiating the atomic fraction of HDA silicon inside and outside a formed SB (i.e., in bulk) versus $\gamma$. After  gradual increase in $c$ during the uniform deformation with increasing $\gamma$, SB forms, in which the atomic fraction of HDA  experiences an surprising abrupt and substantial decrease. After sharp drop, the HDA atomic fraction within the SB stabilizes at a significantly lower value ($\sim 0.25-0.28$ for $9$ GPa and $\sim 0.12-0.15$ for $8$ GPa), which remains practically constant during the further straining. Such a reverse PT to a  steady, lower HDA atomic fraction inside the SB sharply contrasts with the continuously increasing fraction outside the SB. 
 Due to much larger $\gamma$ in the SB, the opposite picture was expected, namely, faster evolution of HDA than in the bulk. Therefore, a  distinct deformation mechanism is expected in the SB.

\begin{figure} [!htbp]
 \centering
 \includegraphics[width=\textwidth]{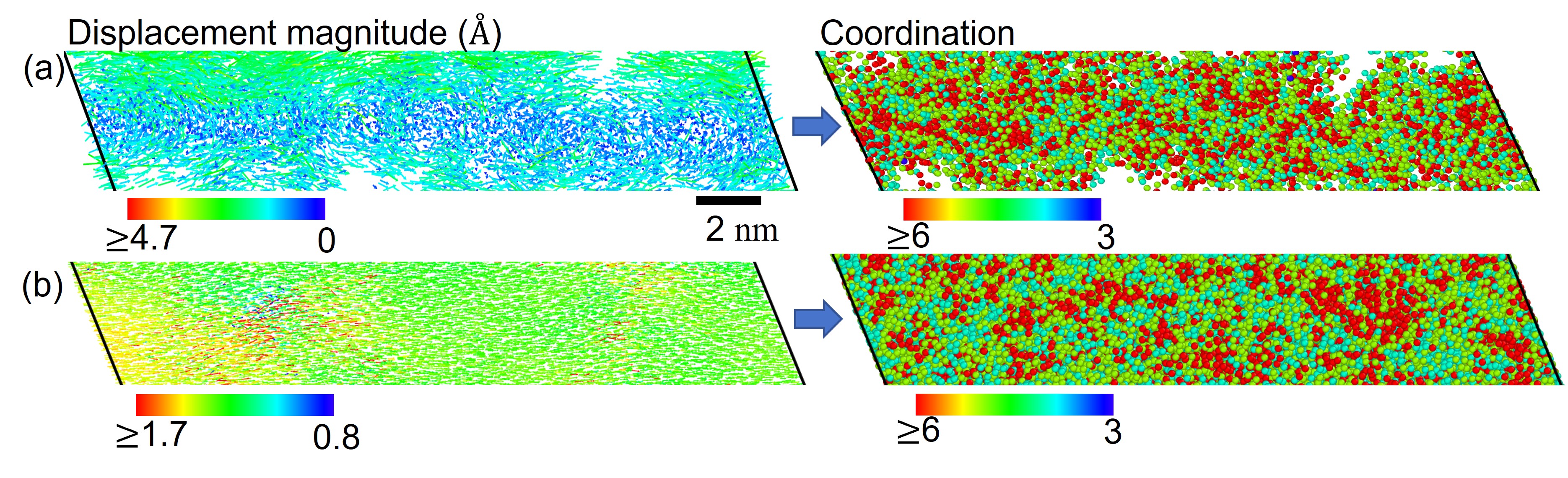}
    \caption{
  {\color{black}  {\bf Atomic displacement fields (left) and coordination distribution (right) within and outside a SB in a-Si at 9 GPa.} (a) Inside of the SB and  (b)  outside of the SB.  Turbulent-like flow with swirls within the SB and perturbed laminar flow outside the band, as well as lack of correlation between atomic displacement and coordination fields  are evident.}
}

  \label{fig:plotS31}
\end{figure}

Left panels in Fig \ref{fig:plotS31} show turbulent-like atomic displacement field within a SB 
and laminar-like shear flow with perturbations outside the band, both similar to those without PT at normal pressure.
The atomic coordination distributions in the right panels  in Fig. \ref{fig:plotS31} do not show any correlation with the displacement fields. Thus, while atomic displacements show a clear difference in the deformation mechanism, it is impossible to connect them directly to the difference in PT mechanism. 

\begin{figure} [!htbp]
 \centering
 \includegraphics[width=0.7\textwidth]{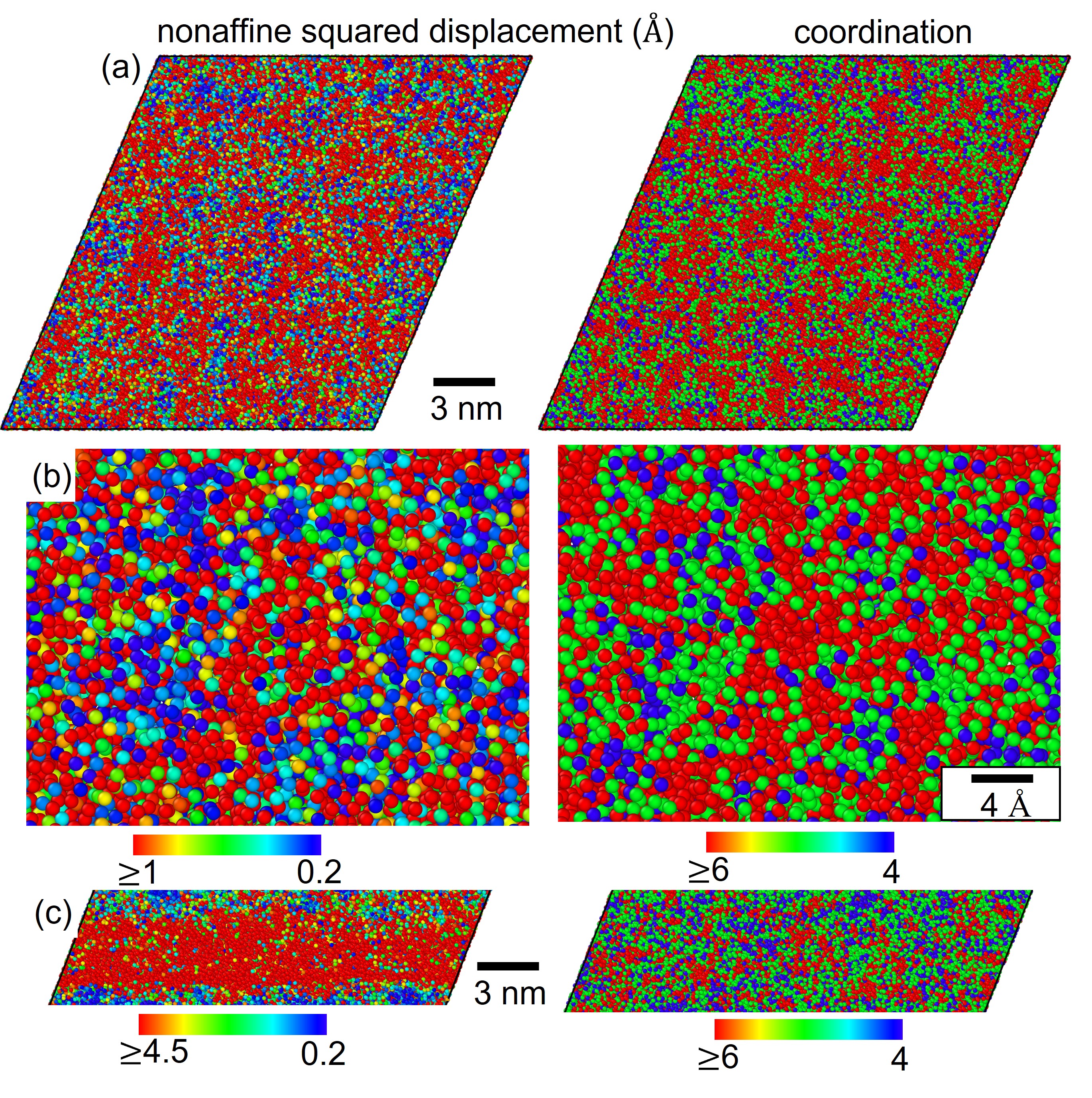}
    \caption{
  {\color{black}  {\bf Atomic mechanism of shear-induced PT in a-Si.}  (a) The spatial distribution of the nonaffine squared displacements $D^2_{NA}$ (left panels) and the corresponding local atomic coordination (right panels) at $9.8$ GPa. (b) A magnified view of a specific local region in (a). There is evident correlation between $D^2_{NA}$ and atomic coordination fields. (c) The spatial distribution of $D^2_{NA}$ within a shear band  and the corresponding local atomic coordination at $8$ GPa. No spatial correlation between these fields exists.
  Outside the band, these distributions are very similar to those in (a). 
}
}
  \label{fig:plotS36}
\end{figure}

Fig. \ref{fig:plotS36} correlates local inelastic deformation with local PT.  The nonaffine squared displacements of atoms $D^2_{NA}$ (left panels in Fig. \ref{fig:plotS36} (a)-(b)) isolate the localized, atomic rearrangements relative to the macroscopic affine (uniform) shear deformation, effectively serving as a metric for local plasticity and structural restructuring. The right panels display the corresponding local atomic coordination distribution for the exact same atomic configurations. 
By comparing both panels, the localized clusters with high  $D^2_{NA}$ clearly overlap with the regions that have transitioned to the HDA phase. Note that outside the band at 8 GPa, these distributions are very similar to those in (a), with a similar correlation. 
This direct spatial correlation provides a clear atomistic picture of the transformation mechanism in bulk. While high hydrostatic pressure provides the thermodynamic driving force favoring the denser HDA state, it is often insufficient to overcome the kinetic energy barrier required for the structural transition. The applied shear deformation acts as the kinetic trigger. It forces the atomic network to undergo localized, nonaffine relative motions—essentially "stirring" the  atomic environment producing local perturbations similar to the thermal fluctuations at high temperatures, and therefore providing  the necessary activation energy.
These rearrangements allow atoms to break existing bonds and locally repack into the more thermodynamically stable, HDA configuration thermodynamically dictated by the elevated pressure. Thus, the  PT is a direct consequence of cumulative, localized, shear-driven atomic rearrangements. 

Note that the LDA-to-HDA transition under hydrostatic pressure in a-Si proceeds via a nucleation and growth mechanism \cite{fan2024microscopic} typical for many first-order PTs in crystalline materials. This traditional pathway is characterized by the formation of a few supercritical nuclei that subsequently expand through the continuous propagation of well-defined phase boundaries, resulting in large, continuous domains of the product phase \cite{fan2024microscopic}.  
In contrast, for shear-induced PT, the HDA phase emerges as a multitude of small, fragmented, and isolated clusters scattered relatively uniformly throughout the LDA matrix, without coalescence and large HDA domains or propagating phase fronts (Fig. \ref{fig:plotS36}, right panels). 
Therefore, the mechanism of shear-induced LDA-to-HDA PT fundamentally differs from  the PT mechanism under hydrostatic loading. 


 In contrast to the deformation-PT processes in bulk,  Fig. \ref{fig:plotS36}(c) demonstrates that during shear banding at $8$ GPa, the non-affine displacements becomes highly concentrated and almost uniform within the SB. However, these localized displacements in the SB show no correspondence with the coordination data, suggesting that the structural evolution within a SB follows a fundamentally different mechanism of that outside the shear band. This qualitatively explains drop in the atomic fraction of HDA within SB in Fig. \ref{fig:plotS28} (b)-(c). However, specific nucleation mechanism of the HDA in a SB and controlling parameters remain to be determined.


\noindent
{ {\bf Mechanism-based kinetic equation for strain-induced LDA$\leftrightarrow$HDA PT} 

 MD simulations revealed that strain-induced LDA$\rightarrow$HDA PT in the bulk (in particular, outside the shear band) occurs via appearance in STZ clusters (nuclei) of HDA phase without their growth and coalescence. Number of STZ zones 
 is scaled with the plastic strain; luck of growth implies that time is not an important parameter. Therefore,  kinetic equation in the form $dc/dq= f(p,c) $, like for strain-induced PT in  crystalline materials \cite{Levitas-PRB-04,Levitas-PMS-26}, should be derived. Here, $q$ is accumulated plastic strain defined as $ \displaystyle{\dot{q} = \left({2}/{3}
{ d}_p^{ij} { d}_p^{ij} \right)^\frac{1}{2}} $ with summation over the repeated indices ($i,j=1,2,3$),
$ { d}_p^{ij} $ are the components of plastic deformation rate.  
Incomplete PT with the stationary atomic fraction of  HDA $c_s$ dependent on pressure and reverse PT during formation of the shear band in Fig. \ref{fig:plotS28} lead  to conclusion that both LDA$\leftrightarrow$HDA PTs occur simultaneously. 
We will take into account that plastic strain $q_l$ in the LDA promotes transformation LDA$\rightarrow$HDA, and simultaneously,  plastic strain $q_h$ in the HDA promotes reverse transformation HDA$\rightarrow$LDA. 
Then we postulate the simplest linear in pressure first-order kinetics for the direct (sign +) and reverse (sign -) PTs  
\bey
 \frac{d  c_+}{d  q_l}=k_d (1-c) (p  -  p_\vep^d) H(p  -  p_\vep^d); \qquad    \frac{d (1- c_-)}{d  q_h}= -\frac{d c_-}{d  q_h}=k_r c (p_\vep^r -p )H(p_\vep^r -p).
\label{nat-7s}
\eey
Here, $k_d$ and $k_r$ are the kinetic coefficients for direct and reverse transformations, respectively; $p_\vep^d$ and $p_\vep^r$ are the minimum pressure for direct and maximum pressure for the reverse strain-induced PTs, respectively; the Heaviside unit step function $H$ ($H(x) = 1$ for $x \geq 0$; $H(x) = 0$ for $x < 0$) guarantees  that the contribution for the direct transformation is nonzero for  $p >  p_\vep^d$ only, while the term for the reverse PT contributes for $p_\vep^r >p$; for $p_\vep^d<p<p_\vep^r$, $H(p) = 1$.

To determine $q$
for each phase, we assume (similar to \cite{Levitas-PRB-04}) 
\bey
  \frac{q_{l}}{q_{h}}  =
\frac{\tau_{yh}}{\tau_{yl}}\;\; {\rm and }\;\;  q  =
(1-c)  q_l  +  c  q_h  \;\; \rightarrow \;\;   q_l   =   q   \frac{\tau_{yh}}{\bar\tau}   ; \quad
q_h   =   q   \frac{\tau_{yl}}{\bar\tau}   ,
\quad  \bar\tau= c   \tau_{yl}   +
\left(1   -   c \right) \tau_{yh}  .
\label{nat-8s}
\eey
Here, $ \tau_{y}$ is the yield strength in shear, and subscripts $h$ and $l$ correspond to HDA and LDA, respectively. 
 Eq.(\ref{nat-8s}) takes into account that  the weaker the phase is, the larger fraction
of the equivalent plastic strain is localized in it. Then, combining equations for direct and reverse transformations with allowing for 
 Eq.(\ref{nat-8s}) results in the final kinetic equation
\bey
 \frac{d  c}{d  q}= \frac{d  c_+}{d  q}+ \frac{d  c_-}{d  q} =\frac{k_d (1-c)}{c\frac{\tau_{yl}}{\tau_{yh}}+(1-c)} (p  -  p_\vep^d) H(p  -  p_\vep^d)-
  \frac{k_r c}{c+(1-c)\frac{\tau_{yh}}{\tau_{yl}}} (p_\vep^r -p )H(p_\vep^r -p).
\label{nat-9s}
\eey
Steady  solution to this equation ($dc/dq=0$) in the pressure range $p_\vep^d \leq p \leq p_\vep^r$ can be presented in the form
\bey
c_s  =  \frac{1}{1  +  M
\left( 1  -  \tilde{p} \right)/\tilde{p}}  ;
\qquad
0 \leq \tilde{p}  = \frac{p  -  p_\vep^d}{p_\vep^r  -  p_\vep^d} \leq 1; \qquad 
M  =  \frac{\tau_{yl}}{\tau_{yh}} \frac{k_r}{k_d} .
\label{nat-10s}
\eey
Fig. \ref{fig:kinetics2}a shows a very good comparison for $c_s(p)$ outside and inside the shear band  and their approximation by analytical Eq.(\ref{nat-10s}).  The following best-fit parameters are obtained: in the bulk,
$M = 7.057$, $p_\vep^d = 2.314$, and  $p_\vep^r = 9.956$; inside the SB, $ M = 9.798$,  $p_\vep^d = 3.484$, and $p_\vep^r = 9.951$. 
Note that because pressure within the band is slightly lower than the outside the band (Fig. \ref{fig:kinetics2}b),  we took this into account in Fig. \ref{fig:kinetics2}a while placing results of MD simulations.

At 9.8 GPa, the yield strength in shear is  $\tau_{yl}\simeq 1.25$ GPa at plateau for $0.25 \leq \gamma \leq 0.80$ when the transformation just started and  $\tau_{yh}\simeq 1.75$ GPa at the second plateau for $1.30 \leq \gamma \leq 1.47$ when the transformation is almost completed (see Fig. \ref{fig:plotS19}(a)). Pressure 9.8 GPa is chosen because of the absence of a SB and presence of almost pure both phases at different $\gamma$. Therefore,  $\frac{\tau_{yl}}{\tau_{yh}}=0.714$; we assume the same values inside the band because $\tau_{yh}$ inside the band cannot be determined due to small atomic fraction of HDA. Based on $M$ and   $\frac{\tau_{yl}}{\tau_{yh}}=0.714$, we obtain $\frac{k_r}{k_d}=9.880$ in the bulk and 13.717 inside the band. 

Generally, this is very unusual situation for strain-induced PTs that the minimum pressure for PT to the high-pressure phase is lower than
the maximum pressure for the reverse PT, and both direct and reverse PTs occur simultaneously. Recent comprehensive review
\cite{Levitas-PMS-26} does not show any such example. While  $\frac{\tau_{yl}}{\tau_{yh}}<1$ promotes direct PT,  $\frac{k_r}{k_d}=9.880$ in bulk and $\frac{k_r}{k_d}=13.717$ in SB strongly promote the reverse PT. Due to large $M$, significant atomic fraction of the HDA is possible at relatively high pressures close to $p_{\varepsilon}^r$, which is surprisingly practically the same within and outside the SB.

To check kinetic  Eq.(\ref{nat-9s}), we recollect that for large  simple shear, the total shear $\gamma$ is the sum of the elastic and plastic parts, $\gamma=\gamma_e+\gamma_p$, and   $q=\gamma_p/\sqrt{3}$. Elastic strain is $\gamma_e=\tau/G$, where $G$ is the shear modulus. It is evident from Figs. \ref{fig:plotS19}a and \ref{fig:plotS18}a that $G \simeq 20$ GPa for all pressures. These equations allow us to determine evolution of $q$ during shearing and express $c$ versus $q$ in  Fig. \ref{fig:kinetics2}c.
The only unknown constant in kinetic  Eq.(\ref{nat-9s}) is $k_d$ (or $k_r$). Fitting it to all MD simulation data for all pressures gives
$k_d=0.1865$ and $k_r=1.8425$ in bulk; due to just few points and low $c$, determination of these constants inside the band is unreliable. Fig. \ref{fig:kinetics2}c shows remarkable correspondence between MD simulations and Eq.(\ref{nat-9s}) for all pressures, unexpected for the first simple analytical model in the field. This means that our assumptions are correct, in particular, about simultaneous occurrence of LDA$\leftrightarrow$HDA PTs. 

The results allow a realistic interpretation of 
why larger plastic shear in a SB causes lower atomic fraction of HDA than in the bulk, i.e., suppresses direct PT. 
In fact, plastic shear does not suppress but promotes both direct and reverse simultaneously. 
However, since $\frac{k_r}{k_d}=9.880$ in the bulk and 13.717 inside the band, we conclude that the transition from STZ in the bulk to turbulent-like flow and vortices promotes reverse PT much stronger than the direct one. 

Note that the effect of shear stresses was neglected due to the following reasons. Contribution to the thermodynamic driving for  PT due to shear is $\tau \gamma_t$, where   $\gamma_t$ is transformation shear. For martensitic PT,  $\gamma_t$ is determined by crystallography and it changes sign for the the reverse PT. For PT in disordered materials, $\gamma_t \sim \tau $ \cite{Levitas-Attariani-SR-13,Levitas-Attariani-JMPS-14,Levitas-IJP-21}
and for the same $\tau$ it occurs in the same direction and with a similar magnitude for direct and reverse PTs. That is why shear stress promotes both direct and reverse PTs approximately equally, and does not contribute to the resultant kinetic Eq.(\ref{nat-9s}).
Excellent agreement between kinetic  Eq.(\ref{nat-9s}) and MD simulations confirms our prediction  of this unique feature of amorphous-amorphous PT.

\begin{figure} [!htbp]
 \centering
 \includegraphics[width=\textwidth]{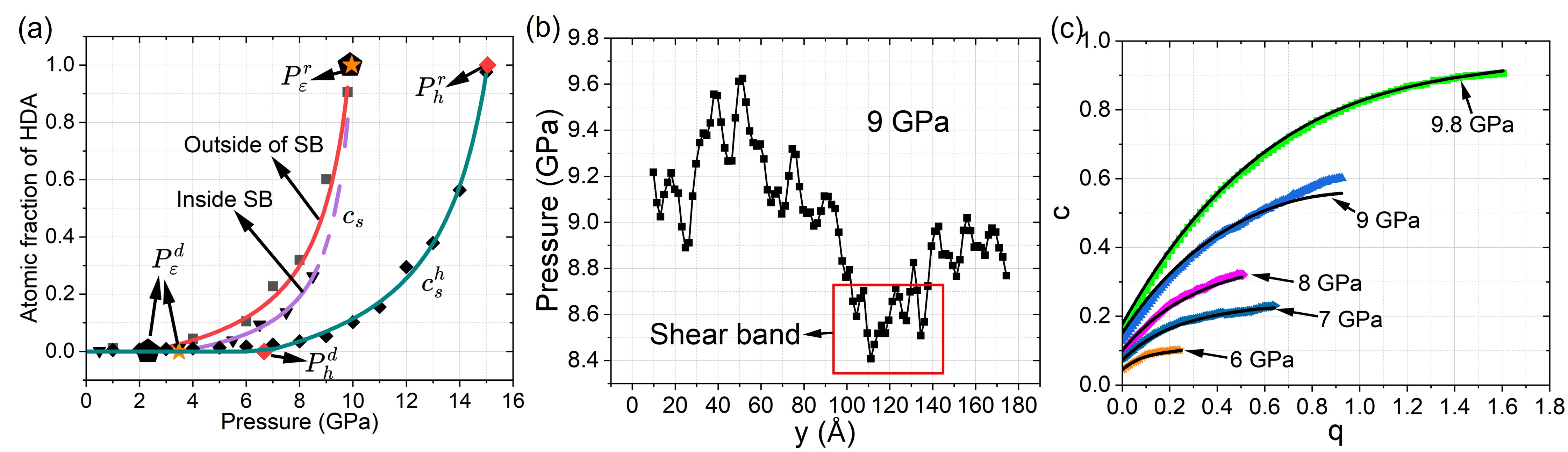}
    \caption{
  {\bf Comparison of MD simulations and analytical theory.}  (a) 
Stationary solution $c_s$ for the atomic fraction of HDA phase versus pressure in bulk (squares) and inside the SB (triangles) for the strain-induced PT and $c_s^h$ for the pressure-induced PT (diamonds), as well as their approximations by analytical solution in Eqs.(\ref{nat-10s}) and (\ref{nat-10sh}) (lines). Dashed line represents extrapolation of data within the SB for high pressures, at which shear bands do not exist. Symbols outside  and inside the shear band
are for slightly different pressures, because pressure within the band is slightly lower than the outside the band (see figure (b)). 
(b) Distribution of pressure averaged over the sample in horizontal (shear) direction along the vertical line, exhibiting the pressure reduction in the shear band. (c) Evolution of the atomic fraction of the  HDA phase $c$ versus accumulated plastic strain $q$ in bulk for various pressures obtained with MD (bold  color lines) and  their approximation by solution of Eq.(\ref{nat-9s}) (thinner black lines). Remarkable correspondence between MD results and analytical kinetic Eqs.(\ref{nat-9s}) and (\ref{nat-10s}) is evident.   
}
  \label{fig:kinetics2}
\end{figure}
Note that for crystalline solids TRIP strain rate tensor is determined by equation $d_{TRIP}^{ij}= \lambda  S^{ij} \vep_0 \dot{c}$ \cite{Fischer-etal-00}, where $S^{ij}$ are the components of the  deviatoric true stress, $\vep_0$ is the volumetric transformation strain, and $\lambda$ is a parameter. Due to the TRIP mechanism described above, for amorphous solid it should generalized as $ d_{TRIP}^{ij}= \lambda   S^{ij} \vep_0 (|\dot{c}_+|+|\dot{c}_-|)$.  
This equation will be checked in the future work.

For hydrostatic loading, the PT occurs via classical nucleation and growth, and plastic strain and STZs are not relevant. Then a similar reasoning but with time $t$ instead of $q$ leads to the kinetic equation 
\bey
 \frac{d  c}{d t}=k_d^h (1-c) (p  -  p_h^d) H(p  -  p_h^d)-
k_r^h c   (p_h^r -p )H(p_h^r -p)
\label{nat-9sh}
\eey
with the stationary  solution in the pressure range $p_h^d \leq p \leq p_h^r$ 
\bey
c_s^h  =  \frac{1}{1  +  M_h
\left( 1  -  \bar{p} \right)/\bar{p}}  ;
\qquad
0 \leq \bar{p}  = \frac{p  -  p_h^d}{p_h^r  -  p_h^d} \leq 1; \qquad 
M_h  =   \frac{k_r^h}{k_d^h} .
\label{nat-10sh}
\eey
Here, $p_h^d$ and $p_h^r$ are the minimum pressure for direct and maximum pressure for the reverse pressure-induced PTs, respectively and $k_d^h$ and $k_r^h$ are the kinetic coefficients for direct and reverse PTs under hydrostatic loading, respectively. 
Fig. \ref{fig:kinetics2}a presents a very good comparison of MD results for $c_s^h(p)$ with analytical Eq.(\ref{nat-10sh}) for 
$ M_h = 5.1207$,  $p_h^d = 6.671$ GPa, and $p_h^r = 15.045$ GPa.
Thus,  plastic straining reduces the minimum  pressure for initiation of LDA$\rightarrow$HDA   PT by 4.36 GPa and the maximum pressure  for the HDA$\rightarrow$LDA PT by 5.10 GPa, and slightly reduces ratio of the kinetic coefficient for the reverse to direct PTs.

}

\section*{Discussion}

In a light of  various actual and potential engineering applications, it is very surprising that strain-induced amorphous-to-amorphous PT under high pressure and its interaction with shear banding were not considered previously, especially for Si.  
Monocrystalline Si remains the foundational material for the global semiconductor and microelectronics industries. The manufacturing of Si-based devices inevitably involves a series of abrasive and ultra-precision machining processes, such as wafer grinding, lapping, diamond turning, and chemical-mechanical polishing. During these processes, the mechanical interaction between the abrasive grits and the Si substrate imposes localized shear strain under high pressure, causing solid-state amorphization and leaving behind a nanoscale a-Si layer on the machined surface \cite{huo2015surface, wang2025systematic}.  Plastic flow, material removal, and surface damage for the subsequent surface treatment are thus largely dictated by the mechanical response of the a-Si layer. 
To reduce subsurface damage, preventing microcrack initiation, and achieving the defect-free surfaces required for next-generation semiconductor manufacturing, special pressure-shear conditions are realized by cutting tool to produce PT as the mechanism of plasticity, thus realizing the regime of ductile machining of Si, Ge, and strong brittle ceramics \cite{Patten-Cherukuri-Yan-section6-04,Goeletal-15}.  
Crystal-amorphous PT is considered under such condition; however, whether to LDA or HDA and whether LDA-to-HDA
PT occurs, and under which conditions, remains completely unclear, in particular, due to reversibility of LDA$\rightarrow$HDA PT. 

Furthermore, a-Si thin films are integral to a wide array of modern technologies, serving as foundational materials for thin-film solar cells, active-matrix displays, and  flexible nanoelectronic. In these advanced applications, the deposited a-Si films are routinely subjected to extreme and complex multiaxial mechanical loadings during operation. Also, during the lithiation-delithiation cycles of a battery, the a-Si host undergoes swelling with volume expansion up to 300$\%$ in the constrained conditions due to a substrate, leading
to the GPa-level stresses and plastic straining \cite{Zhao2012ReactiveFlowLithium,Levitas-Attariani-SR-13}. 
They may cause LDA-to-HDA PT and shear banding.
Shear banding is highly detrimental in engineering applications, as these nanoscale bands act as precursors to crack initiation and brittle fracture, ultimately degrading the structural integrity of the processed silicon wafer, Si-based batteries, and other devices \cite{hedler2004amorphous}. Understanding and controlling the deformation mechanisms, PTs, and SBs  
 are of critical engineering significance for minimizing subsurface damage, suppressing microcrack formation, and achieving ultra-smooth defect-free surfaces. 

Various fundamental results for coupled plastic straining, shear banding and strain-induced amorphous-to-amorphous PT were obtained above. Thus, increasing pressure, despite reduction of the peak and plateau shear strength,   suppresses formation of SB by promoting shear-induced LDA$\rightarrow$HDA PT, reducing the localized shear strain in them and leading to the uniform deformation at 9.8 GPa. This information can be used for controlling surface strain-induced damage, machining processes,  and inducing regime of ductile machining   for crystalline and a-Si and potentially other strong and brittle covalent semiconductors and ceramics 
\cite{huo2015surface, wang2025systematic,Patten-Cherukuri-Yan-section6-04,Goeletal-15}.

Even a concept of strain-induced PTs under high pressure \cite{Levitas-PRB-04,Levitas-PMS-26} was never applied to amorphous-amorphous PTs.  It was found here that plastic straining strongly promotes LDA$\rightarrow$HDA PT, in particular by reducing the pressure for initiation and completion of LDA$\rightarrow$HDA PT from 6.671 and 15.045 GPa under hydrostatic loading to 2.314 and  9.956 GPa, respectively, i.e., 
by 4.36 and 5.10 GPa, respectively. Plastic straining also dramatically changes the LDA$\rightarrow$HDA PT kinetics, which depends on the accumulated plastic strain instead of time and ratio of the yield strength of phases. These are in line with  similar results for strain-induced PTs in crystalline materials 
\cite{Levitas-PRB-04,Blank-Estrin-2014,Levitas-MT-23,Yesudhasetal-NatCom-Si-24,Review-HPT-JAL-24,Review-ceramics-25,Levitas-PMS-26}, 
thus expanding the science of  strain-induced PTs to amorphous materials. 
Also, plastic straining promotes simultaneously  LDA$\leftrightarrow$HDA PTs in different regions until reaching the steady state.
While this possibility was included in the macroscopic kinetic equation for crystalline materials in \cite{Levitas-PRB-04,Levitas-PMS-26}, no examples of such transformations in experiments or atomistic simulations were known before the current paper.   

Atomistic mechanism  of strain-induced LDA$\leftrightarrow$HDA PTs is  fundamentally different from  that for  crystalline materials and from classical nucleation and growth mechanism for pressure-induced PTs in amorphous and crystalline materials. It is also different inside and outside the SB. 
For strain-induced PT in crystalline materials,  the mechanism is related to the nucleation at the tips of plastic strain-generated dislocation pileups  \cite{Levitas-PRB-04,Yesudhasetal-NatCom-Si-24,Levitas-PMS-26}. It was shown here that in bulk the LDA$\rightarrow$HDA PT is triggered by localized nonaffine atomic displacements in STZs within LDA,  producing dispersed  nuclei of HDA without their growth and coalescence. Concurrently, similar nucleation of LDA clusters occurs within HDA phase.
In the SBs, coalesce of STZs transforms into turbulent-like  motion with swirls. This vortex-like structure further evolves with reducing number of swirls due to their coalescence. To some extend   similar  swirl structure was reported only for a metallic glass in  \cite{csopu2017atomic}, but due to order of magnitude smaller shears and  heterogeneous fields along the band in \cite{csopu2017atomic},  reduction in number of swirls through  their coalescence was not found.

Despite the much larger shear and expected atomic fraction of HDA $c$, a surprising sharp drop in $c$ within the SB was discovered.
This leads to the paradoxical conclusion that increasing plastic shear in SB suppresses rather than promotes the PT. 
The paradoxical results is explained by the found  turbulent-like structure that  changes mechanism and kinetics of strain-induced LDA$\leftrightarrow$HDA PT.
Actually, plastic shear does not suppress but promotes both direct and reverse simultaneously. 
However,  transition from STZ in the bulk to the swirls in SBs promote reverse PT much stronger than the direct one. 
While evident  correlation between the nonaffine squared
displacement and the atomic coordination fields in the bulk is revealed, a similar  correlation in a SB is lacking. 
Thus, finding specific atomic  mechanisms that triggers the PT in the SB  is an outstanding basic problem.

Despite the complexity of the physical processes, we developed a simple, mechanism-based analytical kinetic  model that uses accumulated plastic strain $q$ instead of time, simultaneous direct-reverse PTs, pressure-dependent driving forces for both PTs, and ratio of the yield strengths of the phases. Due to two opposite PTs, it possesses a steady-state solution $c_s (p)$, which strongly increases with pressure. All our MD simulation results for different pressures for non-steady processes and steady states are practically perfectly  described by an analytical model, which also led to identification of all material parameters. 
One more surprise was that large shear stresses do not affect the PT kinetics, in contrast to PTs in crystalline materials. It was rationalized that, based on atomic mechanisms, shear stresses simultaneously promote both direct and reverse PTs in different regions, resulting in negligible net effects. 

TRIP due to volumetric transformation strain during LDA$\leftrightarrow$HDA PTs was revealed and applied for explanation of deviation  from the linear elastic behavior at stresses below the instability peak, increasing with increasing pressure. 
  Since both direct and reverse LDA$\leftrightarrow$HDA PTs occur simultaneously, each of them produces TRIP shear of comparable magnitude. This implied  a significant modification in the equation for the TRIP-induced deformation rate as compared to that for crystalline materials. Obtained equation explains how  even for relatively small net $c$, TRIP and stress relaxation can be significant due to multiple back-and-forth events.

Obtained computational results encourage developing an experimental program utilizing in situ x-ray diffraction and Raman studies of plastic strain-induced amorphous-to-amorphous PTs in different materials  using plastic compression in  diamond anvil cell (DAC) \cite{Pandey-Levitas-ActaMat-20,Lin-Levitas-Zr-PT-nanostructure-25}, torsion in static \cite{Blank-Estrin-2014,Pandey-Levitas-ActaMat-20,Yesudhasetal-NatCom-Si-24}  and dynamic \cite{Yesudhas-rDAC-arxiv-25}  rotational DACs, like for crystalline materials.   Obtained kinetic equation is a major part of  macroscopic models 
utilized for the finite-element simulations of the process in a sample deformed and transformed in DAC \cite{Levitas-Zarechnyy-PRB-DAC-10,Feng-Levitas-IJP-BN-DAC-17} and rotational DAC \cite{Levitas-Zarechnyy-PRB-RDAC-10,Feng-Levitas-IJP-BN-RDAC-19}, developing combined experimental-computational studies of these processes with complete determination of $c$ and tensorial  stress  and plastic strain fields (including those that cannot be measured), and identification of all material parameters \cite{Dhar-Levitas-kinetics-24}.   
Then obtained constitutive equations can be used for simulations and optimization of various engineering processes in amorphous materials, like above-mentioned surface treatments, processes in  a-Si thin films,  and lithiation/delitiation in Si anode in lithium-ion batteries.

\section*{Method}

All molecular dynamics (MD) simulations were carried out using the LAMMPS simulation package \cite{plimpton1995fast}. The simulation system contained $74,520$ silicon atoms with length along each direction $l_x  = 17.6\ nm$, $l_y = 17.6\ nm$ and $l_z = 4.9\ nm$ and with periodic boundary conditions applied in all three directions. Interatomic interactions were described using the Gaussian Approximation Potential (GAP) for silicon, a machine-learning-based interatomic potential that provides near first-principles accuracy across a broad range of silicon phases and environments \cite{bartok2018machine, deringer2021origins, fan2024microscopic}. A time step of 1 femtosecond was used throughout all simulations. 

The amorphous silicon structure was generated through a melt–quench process which is the same as in \cite{deringer2021origins}. Starting from crystalline silicon, the system temperature was first raised above the melting point to $1300\ K$ and then quenched down to $1000\ K$ at a cooling rate of $100\ K\ ns^{-1}$, using a Nos$\acute{e}$–Hoover thermostat \cite{evans1985nose}. This quenching step was performed in the NPT ensemble, with all stress components fixed to zero to allow for full relaxation of the system. Following this, the system was rapidly cooled to the target simulation temperatures of $10\ K$. The resulting quenched amorphous phase exhibits a largely fourfold coordinated glassy network, consistent with the established structural features of amorphous silicon \cite{deringer2021origins}.  The atomistic configurations were visualized and analyzed by the OVITO software \cite{stukowski2009visualization}. 

A constant hydrostatic pressure was imposed using a Nos$\acute{e}$-Hoover barostat within the isothermal--isobaric (NPT) ensemble implemented in LAMMPS, enforcing equal normal stresses along the three Cartesian directions during deformation \cite{evans1985nose}. Shear deformation was applied to the equilibrated amorphous configurations using a simple shear protocol at constant shear strain rates of $0.1$, $1.0$ and $10.0\ ns^{-1}$. During deformation, the temperature was maintained at 10 K \cite{evans1985nose}. The mechanical response of the system was monitored via the stress–strain relation, while local atomic configurations were analyzed to characterize the onset of yielding, plastic flow, and the subsequent evolution of shear localization \cite{stukowski2009visualization}. This simulation framework enables a systematic investigation of the effects of strain rate on the shear deformation of amorphous silicon, while leveraging the accuracy of the GAP potential to resolve atomic-scale rearrangements associated with SB formation.

\bibliography{mybib}
\vspace{1\baselineskip} 

{\bf Acknowledgements:}
 HC acknowledges support from the National Natural Science Foundation of China (No.12472097), Natural Science Foundation of Jiangsu Province (No.BK20250151), Postdoctoral Research Foundation of China (No. 2025-M771318), the Research Initiation Fund for Senior Talents of Jiangsu University, China (23JDG040). V. I. L acknowledges ARO (W911NF2420145), NSF (DMR-2246991 and CMMI-2519764), and Iowa State University (Murray Harpole Chair in Engineering).

{\bf Author contributions:} HC, TYL, JYL performed MD simulation and the theoretical analysis; VIL performed theoretical analysis.  HC and VIL  prepared the manuscript.

{\bf Competing interests:}
The authors declare no competing interests.

{\bf Data availability:}
The data supporting this study's findings are available from the corresponding authors upon request.

\end{document}